\documentclass[%
 reprint,
 showpacs,preprintnumbers,
 nofootinbib,
 amsmath,amssymb,
 aps,
]{revtex4-1}

\usepackage{graphicx}
\usepackage{dcolumn}
\usepackage{bm}

\usepackage{amsmath}
\usepackage{ulem}
\usepackage[usenames]{color}

\usepackage{booktabs}

\begin{document}


\title{Spin polarization and color superconductivity in the Nambu--Jona-Lasinio model \\
at finite temperature}

\author{Hiroaki Matsuoka}%
\email{b16d6a01@s.kochi-u.ac.jp}
\affiliation{Graduate School of Integrated Arts and Science, Kochi University, Kochi 780-8520, Japan}

\author{Yasuhiko Tsue}
\email{tsue@kochi-u.ac.jp}
\affiliation{Physics Division, Faculty of Science, Kochi University, Kochi 780-8520, Japan}

\author{Jo\~{a}o da Provid\^{e}ncia}%
\author{Constan\c{c}a Provid\^{e}ncia}
\affiliation{CFisUC, Departamento de F\'{i}sica, Universidade de Coimbra, 3004-516 Coimbra, Portugal}

\author{Masatoshi Yamamura}
\affiliation{Department of Pure and Applied Physics, Faculty of Engineering Science, Kansai University,
Suita 564-8680, Japan}


\date{\today}

\begin{abstract}
We investigate the possible existence of spin polarization and color superconductivity in the Nambu--Jona-Lasinio model with a tensor-type interaction at finite density and temperature.
The thermodynamic potential is calculated by the functional integral method.
Numerical results indicate that at low temperature and quark chemical potential the chiral condensed phase exists, and at intermediate chemical potential the color superconducting phase appears.
In addition, depending on the magnitude of the tensor coupling, at large chemical potential and low temperature, a color superconducting phase and a spin polarized phase may coexist while at intermediate temperatures only the spin polarized phase occurs.
\end{abstract}



\maketitle


\section{Introduction}
One of the most interesting topics in high energy physics is to clarify the phase structure of quantum chromodynamics (QCD).
The phase where we live is called hadronic phase.
A remarkable feature in this phase is that quarks and gluons are confined.
At high temperature, the quark-gluon plasma (QGP) phase is realized.
Quarks and gluons are not confined in the QGP phase.
This phase has been confirmed by high energy accelerator experiments, for example, the relativistic heavy ion collider experiment.
Moreover, more powerful experiments have been conducted by the large hadron collider.

On the other hand, the phase structure at low temperature and large chemical potential has not been understood very well.
Since presently, it is still not possible to test the low temperature and large chemical potential regime in the laboratory, and we cannot use the lattice simulation method because of the ``sign problem,'' the features of this region of the QCD phase diagram are still uncertain.
In order to investigate the nature at such conditions the Nambu--Jone-Lasinio (NJL) model \cite{Nambu:1961_1,Nambu:1961_2,Klevansky:1992,Hatsuda:1994} has been used.
It has been considered that a color superconducting phase \cite{Alford:2001,Buballa:2002,Buballa:2005}, which may occur in different forms such as the two-flavor color superconducting (2SC) phase or the color flavor-locked (CFL) phase, may be realized under these conditions.
This phase may appear inside compact stars, such as neutron stars.
Compact stars are very dense astrophysical objects which may have very strong magnetic fields \cite{Harding:2006}.
However, the mechanism that explains the generation of such strong magnetic fields is still not completely understood.
In particular, the phase structure at large chemical potential and the possible existence of a spin polarized phase should be investigated.

The possible existence of a quark ferromagnetic phase has been discussed with one-gluon-exchange interaction in Ref.\cite{Tatsumi:2000}.
Moreover, the possibility that spins of quarks may polarize at large chemical potential has been studied with axial vector-type interaction in Refs. \cite{Tatsumi:2003,Tatsumi:2004,Nakano:2003}.
In Ref. \cite{Maedan:2007}, a vector-type interaction which respects chiral symmetry has been introduced in the NJL model and it has been shown that spin polarization could occur if the chemical potential is within a narrow range of values.
The relationship between the vector-type interaction and the 2SC has been discussed and it has been indicated that chiral condensed phase and 2SC phase may coexist if the contribution from the vector-type interactions considered in Ref. \cite{Kitazawa:2002}.

Although a term for spin polarization can be derived from the vector-type interaction, we pay attention to a tensor-type interaction, which, of course, respects chiral symmetry.
The interaction has been introduced in Refs. \cite{Bohr:2012,Tsue:2012} and a spin polarization term can be derived from it.
Note that the spin polarization term from the tensor-type interaction is not identical to that from the vector-type interaction, and the term from the tensor-type interaction can be interpreted as an anomalous magnetic moment induced dynamically according to Ref. \cite{Ferrer:2014}.
In Refs. \cite{Tsue:2013,Tsue:2015_1} the relationship between the spin polarization and the color superconductivity has been investigated at zero temperature, and in our preceding paper \cite{Matsuoka:2016} the discussion has been extended to finite temperature.
According to our preceding paper, the chiral condensed phase and the spin polarized phase do not coexist, and the order of the phase transition from the spin not-polarized phase to the spin polarized phase is second order.
The effect from an external magnetic field on the spin polarization has been studied in Ref. \cite{Tsue:2015_2}, and it has been shown that ferromagnetism may occur if we assume an anomalous magnetic moment for the quarks.
In the present work, we investigate the possible existence of spin polarization and color superconductivity in the Nambu--Jona-Lasinio model with a tensor-type interaction at finite density and temperature.

In the following section we introduce the NJL model with the tensor-type interaction and calculate the thermodynamic potential by the functional integral method.
In Sec. \ref{sec3} we evaluate the thermodynamic potential numerically.
The last section is devoted to the conclusions and remarks.
The tensor-type interaction can be derived from the scalar-interaction channel in the NJL model, however, we treat the coupling constant for the tensor-type interaction as a free parameter.
We use the gamma matrices in the Dirac representation, and adopt the metric tensor: $g^{\mu \nu}=\text{diag}(1,-1,-1,-1)$.

\section{NJL model with tensor-type interaction and quark pairing interaction}
\label{sec2}
We start from the NJL model with the tensor-type interaction and quark pairing interaction at finite quark chemical potential $\mu$.
The Lagrangian density with flavor $SU(2)$ and color $SU(3)$ symmetry at chiral limit can be expressed as
\begin{align*}
\mathcal{L}_{\text{total}} &=\mathcal{L}_{\text{NJL}}+\mathcal{L}_T+\mathcal{L}_C
+\mu \bar{\psi} \gamma^0 \psi, \\
\mathcal{L}_{\text{NJL}} &= \bar{\psi} i \gamma^\mu \partial_\mu \psi + 
G_S \left \{ (\bar{\psi} \psi)^2+(\bar{\psi} i \gamma^5 \vec{\tau} \psi)^2 \right \}, \\
\mathcal{L}_T &= -\frac{G_T}{4} \Big \{ (\bar{\psi} \gamma^\mu \gamma^\nu \vec{\tau} \psi) \cdot 
(\bar{\psi} \gamma_\mu \gamma_\nu \vec{\tau} \psi) \notag \\
&\quad +(\bar{\psi} i \gamma^5 \gamma^\mu \gamma^\nu \psi)
(\bar{\psi} i \gamma^5 \gamma_\mu \gamma_\nu \psi) \Big \}, \\
\mathcal{L}_C &= \frac{G_C}{2} \sum_{A=2,5,7} \Big\{ (\bar{\psi} i \gamma^5 \tau_y \lambda_A \psi^C)
(\bar{\psi}^C i \gamma^5 \tau_y \lambda_A \psi) \notag \\
& \quad + (\bar{\psi} \tau_y \lambda_A \psi^C)(\bar{\psi}^C \tau_y \lambda_A \psi) \Big\},
\end{align*}
where $\tau_i \; (i=x,y,z)$ is Pauli matrix for flavor space and $\lambda_A \; (A=2,5,7)$ is
Gell-Mann matrix for color space.
The superscript $C$ means charge conjugate.
Note that $\mathcal{L}_T$ and $\mathcal{L}_C$ are the tensor-type interaction term and the quark pairing interaction term, respectively.
In order to study the system at finite density, we introduce quark chemical potential $\mu$.
Here we use the same value for the quark chemical potential for up- and down-quarks.
The spin polarization term appears from $\mathcal{L}_T$ when $\mu=1, \; \nu=2$ or $\mu=2, \; \nu=1$
as follows:
\begin{equation*}
\varSigma_z=-i \gamma^1 \gamma^2=
\begin{pmatrix}
\sigma_z & 0 \\ 0 & \sigma_z
\end{pmatrix},
\end{equation*}
where $\sigma_z$ is the third component of Pauli matrices.

Since we ignore the collective excitations on the realized vacuum in this paper,
the Lagrangian density that we consider here is as follows:
\begin{align}
\mathcal{L}&=\bar{\psi} i \gamma^\mu \partial_\mu \psi + G_S (\bar{\psi}\psi)^2 + 
\frac{G_T}{2} (\bar{\psi} \varSigma_z \tau_z \psi)^2 \notag \\
&\quad -\frac{G_C}{2} (\bar{\psi} \gamma^5 \tau_y \lambda_2 \psi^C)
(\bar{\psi}^C \gamma^5 \tau_y \lambda_2 \psi) + \mu \bar{\psi} \gamma^0 \psi.
\end{align}
We will calculate the thermodynamic potential by the functional integral method.
Let us introduce the generating functional $Z$:
\begin{align}
Z \propto \int & \mathcal{D} \bar{\psi} \mathcal{D} \psi \exp \Big[ i \int d^4 x 
\Big\{ \bar{\psi} i \gamma^\mu \partial_\mu \psi + G_S (\bar{\psi}\psi)^2 \notag \\
&+\frac{G_T}{2}(\bar{\psi} \varSigma_z \tau_z \psi)^2
-\frac{G_C}{2} (\bar{\psi} \gamma^5 \tau_y \lambda_2 \psi^C)
(\bar{\psi}^C \gamma^5 \tau_y \lambda_2 \psi) \notag \\
& + \mu \bar{\psi} \gamma^0 \psi \Big\} \Big].
\end{align}
We introduce the auxiliary fields in order to perform functional integral with respect to quark fields.
The auxiliary fields that we introduce here are as follows:
\begin{align}
1 &= \int \mathcal{D} M \exp \Big[ -i \int d^4 x 
\left\{ M/2 + G_S(\bar{\psi}\psi) \right\} \notag \\
& \qquad \qquad \times G_{S}^{-1} \left\{ M/2 + G_S (\bar{\psi} \psi) \right\} \Big], \notag \\
1 &= \int \mathcal{D} F \exp \Big[ -\frac{i}{2} \int d^4 x \left\{ 
F+G_T(\bar{\psi} \varSigma_z \tau_z \psi) \right\} \notag \\
& \qquad \qquad \times G_{T}^{-1} \left\{ F+G_T(\bar{\psi} \varSigma_z \tau_z \psi) \right\} \Big], \notag \\
1 &= \int \mathcal{D} \varDelta^\dagger \mathcal{D} \varDelta
\exp \Big[ -\frac{i}{2} \int d^4 x 
\left\{ \varDelta^\dagger + G_C(\bar{\psi}^C \gamma^5 \tau_y \lambda_2 \psi)^\dagger \right\} \notag \\
&\qquad \qquad \times G_C^{-1} \left\{ \varDelta + G_C(\bar{\psi}^C \gamma^5 \tau_y \lambda_2 \psi) \right\} \Big]
\notag \\
&= \int \mathcal{D} \varDelta^\dagger \mathcal{D} \varDelta
\exp \Big[ -\frac{i}{2} \int d^4 x 
\left\{ \varDelta^\dagger - G_C(\bar{\psi} \gamma^5 \tau_y \lambda_2 \psi^C) \right\} \notag \\
&\qquad \qquad \times G_C^{-1} \left\{ \varDelta + G_C(\bar{\psi}^C \gamma^5 \tau_y \lambda_2 \psi) 
\right\} \Big].
\notag
\end{align}
Inserting the above auxiliary fields into the generating functional, we obtain
\begin{align}
Z &\propto \int \mathcal{D} \bar{\psi} \mathcal{D} \psi \mathcal{D} M \mathcal{D} F \mathcal{D} 
\varDelta^\dagger \mathcal{D} \varDelta \notag \\
& \exp \Big[ i \int d^4 x \Big\{ \bar{\psi}(i \gamma^\mu \partial_\mu - M)\psi
-\bar{\psi} F \varSigma_z \tau_z \psi \notag \\
& \qquad -\frac{1}{2} \varDelta^\dagger \bar{\psi}^C \gamma^5 \tau_y \lambda_2 \psi 
+ \frac{1}{2} \varDelta \bar{\psi} \gamma^5 \tau_y \lambda_2 \psi^C + \mu \bar{\psi} \gamma^0 \psi
\notag \\
& \qquad -\frac{M^2}{4G_S}-\frac{F^2}{2G_T}-\frac{|\varDelta|^2}{2G_C}\Big\} \Big].
\end{align}
In order to transform the above expression into bilinear form for quark fields, we decompose it into
\begin{align*}
Z & \propto \int \mathcal{D} \bar{\psi} \mathcal{D} \psi \mathcal{D} M \mathcal{D} F \mathcal{D} 
\varDelta^\dagger \mathcal{D} \varDelta \notag \\
& \exp \Big[ \frac{i}{2} \int d^4 x \Big\{ 
\bar{\psi}(i \gamma^\mu \partial_\mu -M)\psi - 
\bar{\psi}^C(i \gamma^\mu \overleftarrow{\partial}_\mu-M)\psi^C \notag \\
&\qquad -\bar{\psi} F \varSigma_z \tau_z \psi + \bar{\psi}^C F \varSigma_z \tau_z \psi^C \\
& \qquad -\varDelta^\dagger \bar{\psi}^C \gamma^5 \tau_y \lambda_2 \psi
+ \varDelta \bar{\psi} \gamma^5 \tau_y \lambda_2 \psi^C \notag \\
&\qquad +\mu \bar{\psi} \gamma^0 \psi -\mu \bar{\psi}^C \gamma^0 \psi^C \notag  \Big\} \Big] \notag \\
&\times \exp \Big[ -i \int d^4 x \Big\{ 
\frac{M^2}{4G_S}+\frac{F^2}{2G_T}+\frac{|\varDelta|^2}{2G_C} \Big\} \Big]. \notag
\end{align*}
Let us define the Nambu spinors:
\begin{equation}
\varPsi(x) := \frac{1}{\sqrt{2}}
\begin{pmatrix}
\psi(x) \\ \psi^C(x)
\end{pmatrix},\quad
\bar{\varPsi}(x) := \frac{1}{\sqrt{2}}
\begin{pmatrix}
\bar{\psi}(x) & \bar{\psi}^C(x)
\end{pmatrix}.
\end{equation}
Using these spinors, we can rewrite the generating functional into bilinear form of quarks as follows:
\begin{align}
Z &\propto \int \mathcal{D} \bar{\psi} \mathcal{D} \psi \mathcal{D} M \mathcal{D} F 
\mathcal{D} \varDelta^\dagger \mathcal{D} \varDelta \notag \\
& \exp \left[ i \int d^4 x \left\{ \bar{\varPsi}(x) S^{-1}(x) \varPsi(x) 
-\frac{M^2}{4G_S}-\frac{F^2}{2G_T}-\frac{|\varDelta|^2}{2G_C} \right\} \right].
\end{align}
Here we define the inverse propagator in position space:
\begin{equation}
S^{-1}(x) := 
\begin{pmatrix}
s_{11} & s_{12} \\ s_{21}& s_{22}
\end{pmatrix},
\end{equation}
and
\begin{align*}
&s_{11} := (i \gamma^\mu \partial_\mu -M+\mu \gamma^0) \mathbf{1}_F \mathbf{1}_C -F \varSigma_z \tau_z \mathbf{1}_C, \\
&s_{12} := \varDelta \gamma^5 \tau_y \lambda_2, \\
&s_{21} := -\varDelta^\dagger \gamma^5 \tau_y \lambda_2, \\
&s_{22} := (-i \gamma^\mu \overleftarrow{\partial}_\mu -M-\mu \gamma^0) \mathbf{1}_F \mathbf{1}_C +F \varSigma_z \tau_z \mathbf{1}_C,
\end{align*}
where $\mathbf{1}_F$ and $\mathbf{1}_C$ are the unit matrices for flavor space and color space, respectively.
We can integrate $Z$ with respect to $\bar{\psi}$ and $\psi$, then we obtain
\begin{align*}
Z &\propto \int \mathcal{D} M \mathcal{D} F \mathcal{D} \varDelta^\dagger \mathcal{D} \varDelta \\
&\exp \Big[ \frac{1}{2} \log \text{Det} \, S^{-1}(x) 
-i \int d^4x \Big\{ \frac{M^2}{4G_S}+\frac{F^2}{2G_T}+\frac{|\varDelta|^2}{2G_C} \Big\} \Big],
\end{align*}
where $\text{Det}$ means functional determinant over position space, the Nambu space, gamma matrices, flavor and color space.
To compute $\text{Det} \, S^{-1}(x)$ we move to momentum space.
The generating functional in momentum space becomes
\begin{align}
&Z \propto \int \mathcal{D} M \mathcal{D} F \mathcal{D} \varDelta^\dagger \mathcal{D} \varDelta 
\notag \\
& \exp \Big[ \frac{1}{2} \int d^4 x \int \frac{d^4 p}{(2\pi)^4} 
\log \text{det} \, \tilde{S}^{-1}(p) \notag \\
&\qquad -i \int d^4 x \Big\{
\frac{M^2}{4G_S}+\frac{F^2}{2G_T}+\frac{|\varDelta|^2}{2G_C} \Big\} \Big],
\end{align}
where $\text{det}$ is for the Nambu space, gamma matrices, flavor space and color space, and 
we introduce the inverse propagator in momentum space as follows:
\begin{equation}
\tilde{S}^{-1}(p) := 
\begin{pmatrix}
\tilde{s}_{11} & \tilde{s}_{12} \\ \tilde{s}_{21}& \tilde{s}_{22} 
\end{pmatrix},
\end{equation}
where
\begin{align*}
&\tilde{s}_{11} := (\not p -M+\mu \gamma^0) \mathbf{1}_F \mathbf{1}_C -F \varSigma_z \tau_z \mathbf{1}_C, \\
&\tilde{s}_{12} := \varDelta \gamma^5 \tau_y \lambda_2, \\
&\tilde{s}_{21} := -\varDelta^\dagger \gamma^5 \tau_y \lambda_2, \\
&\tilde{s}_{22} := (\not p -M-\mu \gamma^0) \mathbf{1}_F \mathbf{1}_C +F \varSigma_z \tau_z \mathbf{1}_C.
\end{align*}

Calculating $\text{det} \, \tilde{S}^{-1}(p) = 0$ for $p_0$ gives us single-particle energies for quasiparticles.
Thus, for simplicity, we introduce two kinds of single-particle energies:
\begin{align}
\varepsilon^{(\alpha,\beta)} &:= 
\sqrt{p_z^2+\left( \sqrt{p_x^2+p_y^2+M^2}+\alpha F \right)^2}+\beta \mu \\
E^{(\alpha,\beta)} &:=
\sqrt{\left( \varepsilon^{(\alpha,\beta)} \right)^2+|\varDelta|^2},
\end{align}
where $\alpha = \pm$ and $\beta = \pm$.

After calculating $\text{det} \, \tilde{S}^{-1}(p)$, we get
\begin{align*}
& \log \text{det} \, \tilde{S}^{-1}(p) \\
&= \log \Big[ \Big\{ \prod_{\alpha,\beta=\pm} 
\big( p_0 -\varepsilon^{(\alpha,\beta)} \big) \big( p_0 +\varepsilon^{(\alpha,\beta)} \big) \Big\}^{N_F} \\
& \qquad \times \Big\{ \prod_{\alpha,\beta=\pm}
\big( p_0 -E^{(\alpha,\beta)} \big) \big( p_0 +E^{(\alpha,\beta)} \big) \Big\}^{2N_F} \Big],
\end{align*}
where $N_F$ means the number of flavor (in this case $N_F =2$).
In order to calculate the Matsubara summation later, we differentiate and integrate the above expression
with respect to single-particle energies:
\begin{align*}
& \log \text{det} \, \tilde{S}^{-1}(p) \\
&= N_F \sum_{\alpha,\beta=\pm} \Big\{ 
\int d \varepsilon^{(\alpha,\beta)} \Big( \frac{1}{p_0+\varepsilon^{(\alpha,\beta)}} 
- \frac{1}{p_0 -\varepsilon^{(\alpha,\beta)}} \Big) \\
& \qquad +2 \int d E^{(\alpha,\beta)} 
\Big( \frac{1}{p_0+E^{(\alpha,\beta)}} -\frac{1}{p_0 -E^{(\alpha,\beta)}} \Big)\Big\}.
\end{align*}
To discuss the system at finite temperature, we use the following substitution:
\begin{equation*}
\int \frac{d^4p}{i(2\pi)^4} f(p_0,\vec{p}) \rightarrow
T \sum_{n=-\infty}^{\infty} \int \frac{d^3 p}{(2\pi)^3} f(i \omega_n,\vec{p}),
\end{equation*}
where $T$ is temperature and $\omega_n := (2n+1)\pi T,\,(n \in \mathbb{Z})$ is the Matsubara frequency for fermion.
Using this substitution, the component of the generating functional becomes
\begin{align*}
&\frac{1}{2} \int \frac{d^4 p}{(2\pi)^4} \log \text{det} \, \tilde{S}^{-1}(p) \\
& \rightarrow \frac{i T}{2} N_F \sum_{n=-\infty}^{\infty} \int \frac{d^3 p}{(2\pi)^3} 
\sum_{\alpha,\beta=\pm} \\
&\Big\{ \int d \varepsilon^{(\alpha,\beta)} \Big( 
\frac{1}{i \omega_n +\varepsilon^{(\alpha,\beta)}} - \frac{1}{i \omega_n - \varepsilon^{(\alpha,\beta)}} 
\Big) \\
&\quad +2 \int d E^{(\alpha,\beta)} \Big( \frac{1}{i \omega_n + E^{(\alpha,\beta)}} 
- \frac{1}{i \omega_n - E^{(\alpha,\beta)}} \Big) \Big\}.
\end{align*}
We can calculate the Matsubara summation with the following formula:
\begin{equation*}
\lim_{\eta \to +0} T \sum_{n=-\infty}^{\infty} \frac{e^{i \omega_n \eta}}{i \omega_n - x}
= \lim_{\eta \to +0} \frac{e^{i \omega_n \eta}}{1+e^{x/T}}
= \frac{1}{1+e^{x/T}}.
\end{equation*}
Using the above formula and integrating with respect to energies, the above expression becomes
\begin{align*}
\frac{i}{2} N_F &\int \frac{d^3 p}{(2\pi)^3} \sum_{\alpha,\beta} \Big[ 
\varepsilon^{(\alpha,\beta)} + 2T \log \Big( 1+\exp [-\varepsilon^{(\alpha,\beta)}/T] \Big) \\
&+ 2 \Big\{ E^{(\alpha,\beta)} + 2T \log \Big( 1+\exp [-E^{(\alpha,\beta)}/T] \Big) \Big\} \Big] \\
&+ \text{const},
\end{align*}
where ``const'' is a constant of integration.\footnote{Here we interchange $\sum_{n=-\infty}^{\infty}$ and
$\int d^3 p$ to calculate the Matsubara summation.
This interchange is not correct mathematically, however, the resulting generating functional is correct.
This technique has been used in Ref. \cite{Bellac}}
Since constant terms do not contribute to thermodynamics, we ignore them.
Substituting the result into the generating functional, $Z$ becomes
\begin{align}
Z &\propto \int \mathcal{D} M \mathcal{D} F \mathcal{D} \varDelta^\dagger \mathcal{D} \varDelta \notag \\
& \exp \Big[ i N_F \int d^4 x \int \frac{d^3 p}{(2\pi)^3} \sum_{\alpha,\beta} \Big\{ 
\frac{\varepsilon^{(\alpha,\beta)}}{2} + E^{(\alpha,\beta)} \notag \\
&\quad + T \log \Big( 1+\exp[-\varepsilon^{(\alpha,\beta)}/T] \Big) \Big( 
1+\exp[-E^{(\alpha,\beta)}/T] \Big)^2 \Big\} \Big] \notag \\
& \times \exp \Big[ -i \int d^4 x \Big\{ \frac{M^2}{4G_S}+\frac{F^2}{2G_T}+\frac{|\varDelta|^2}{2G_C} \Big\} \Big].
\end{align}
In one-loop approximation we obtain the thermodynamic potential as follows:
\begin{align}
&V(M,F, | \varDelta |,\mu, T) \notag \\
&=-N_F \int \frac{d^3 p}{(2\pi)^3} \sum_{\alpha,\beta} \Big\{ 
\frac{\varepsilon^{(\alpha,\beta)}}{2} + E^{(\alpha,\beta)} \notag \\
&\quad + T \log \Big( 1+\exp[-\varepsilon^{(\alpha,\beta)}/T] \Big) 
\Big( 1+\exp[-E^{(\alpha,\beta)}/T] \Big)^2 \Big\} \notag \\
&\quad + \frac{M^2}{4G_S}+\frac{F^2}{2G_T}+\frac{|\varDelta|^2}{2G_C}.
\end{align}

\section{Numerical results}
\label{sec3}

\begin{table}[b]
\caption{Parameter sets used in the present study}
\label{parameter}
	\begin{center}
		\begin{tabular}{ccccc}
		\hline
		Model & $\varLambda/\text{GeV}$ & $G_S/\text{GeV}^{-2}$ & $G_T$ & $G_C/ \text{GeV}^{-2}$ \\
		\hline
		\hline
		GT0 & $0.631 $ & $5.5 $ & $0 $ & $6.6$ \\ \hline
		GT2 & $0.631$ & $5.5$ & $2 G_S$ & $6.6$ \\ \hline
		GT2.6 & $0.631$ & $5.5$ & $2.6 G_S$ & $6.6$ \\ \hline
		\end{tabular}
	\end{center}
\end{table}

\begin{figure*}
	\centering
	\includegraphics[width=2\columnwidth]{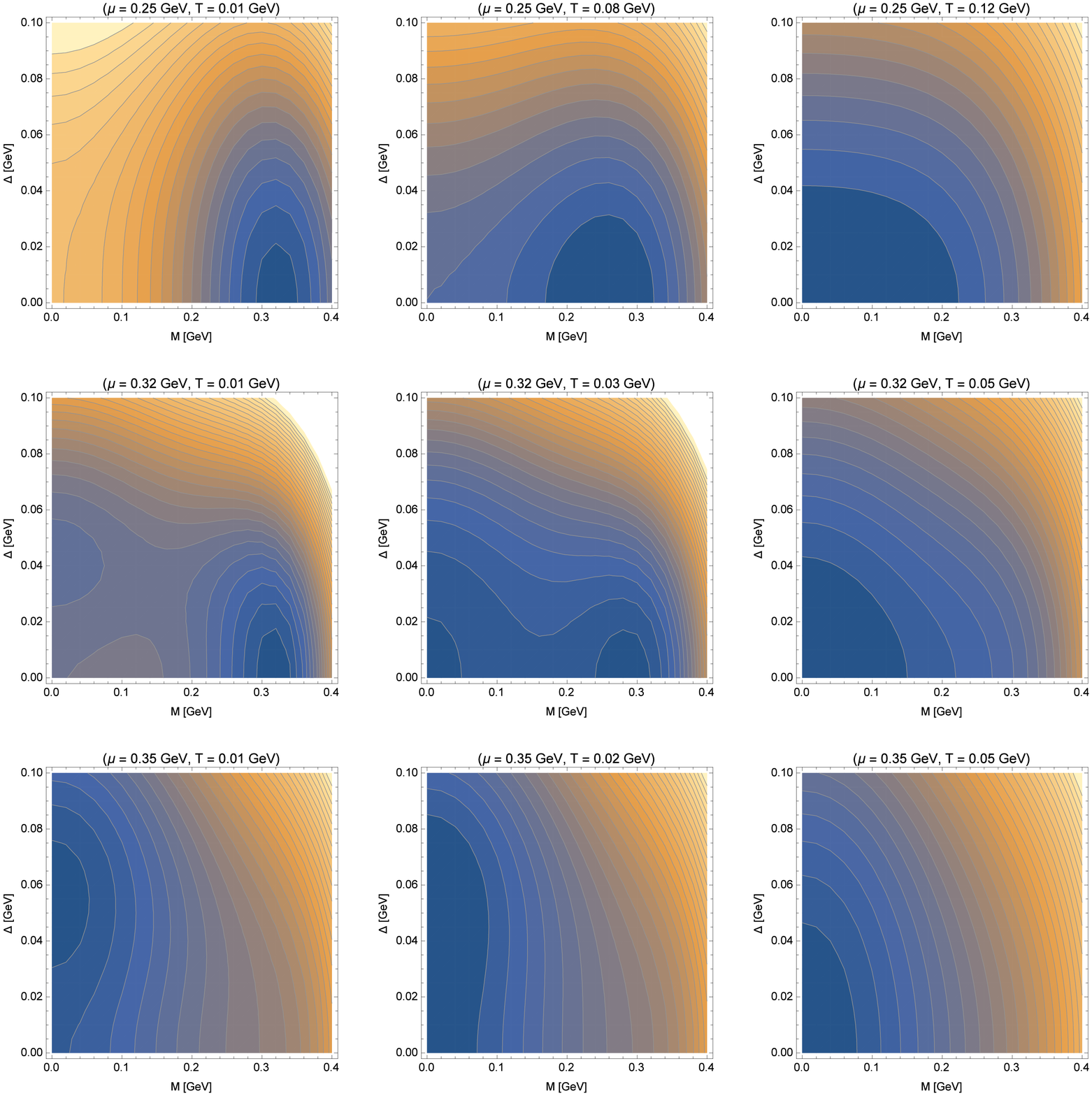}
	\caption{The figures show the contour maps of the thermodynamic potential for model GT0 with several values of the chemical potential, $\mu$, and temperature, $T$.
	The horizontal and vertical axes represent the order parameters $M$ and $\varDelta$ for the chiral condensate and the color superconducting gap, respectively.
	The darker color represents lower values of the thermodynamic potential.}
	\label{map1}
\end{figure*}

In this section we calculate the thermodynamic potential numerically.
To do this we use parameters in Table \ref{parameter}.
Since the NJL model is not a renormalizable theory, we adopt a three-momentum cutoff parameter, 
$\varLambda$.
The values of $G_S$ and $\varLambda$ are determined to reproduce the chiral condensate or dynamical quark mass 
and the pion decay constant in the vacuum. 

We consider that the tensor-type interaction should be derived 
from a two-gluon exchange interaction in QCD \cite{Matsuoka:2016}. 
However, since the NJL model cannot be derived from the QCD Lagrangian directly, 
we therefore adopt $G_T$ as a free parameter in this model. 
If we assume that the tensor-type interaction term is derived by the Fierz transformation of the scalar and
pseudoscalar channels in the NJL model, as in \cite{Nambu:1961_2} or \cite{Blin:1988}, the tensor coupling satisfies the relation $G_T = -2/13 \; G_S$. 
On the other hand, the value of $G_T$ and $G_S$ may be considered independently as in \cite{Jaminon:1998, Jaminon:2002} or \cite{Battistel:2016} and determined from the vacuum mesonic properties.
In \cite{Jaminon:1998, Jaminon:2002}, the scalar and tensor terms and three-dimensional cutoff were used to describe the pion and $\rho$-meson, and the relation $G_S \sim -1.2 G_S$ was obtained.
Recently the authors of \cite{Battistel:2016} have calculated within an extended $SU(2)$ NJL model including vector, axial vector and tensor interactions several meson masses, meson-quark coupling constants and corresponding decay constants within a Hartree plus random phase approximation.
They have considered both positive and negative tensor couplings, and in both cases could describe the mesonic phenomenology.
Although their model is different from ours because of the regularization procedure used and the inclusion of a vector contribution besides the scalar and the tensor ones, they obtained a $G_T$ 4 times larger than $G_S$ when the  same sign of $G_S$, as in our work, is considered.
We must take a positive coupling constant in order to get a spin condensate.
However, we should point out that since we discuss the system at finite density, taking the couplings that have been obtained from the vacuum properties may not be adequate.
Thus, we treat $G_T$ as a free parameter. 

Further, the value of $G_C$ has been used in Ref.\cite{Kitazawa:2002}. 
The value has been taken in order to reproduce the phase diagram in Ref.\cite{Berges:1999}.

\subsection{Chiral condensate versus color superconducting gap}

\begin{figure*}
	\centering
	\includegraphics[width=2\columnwidth]{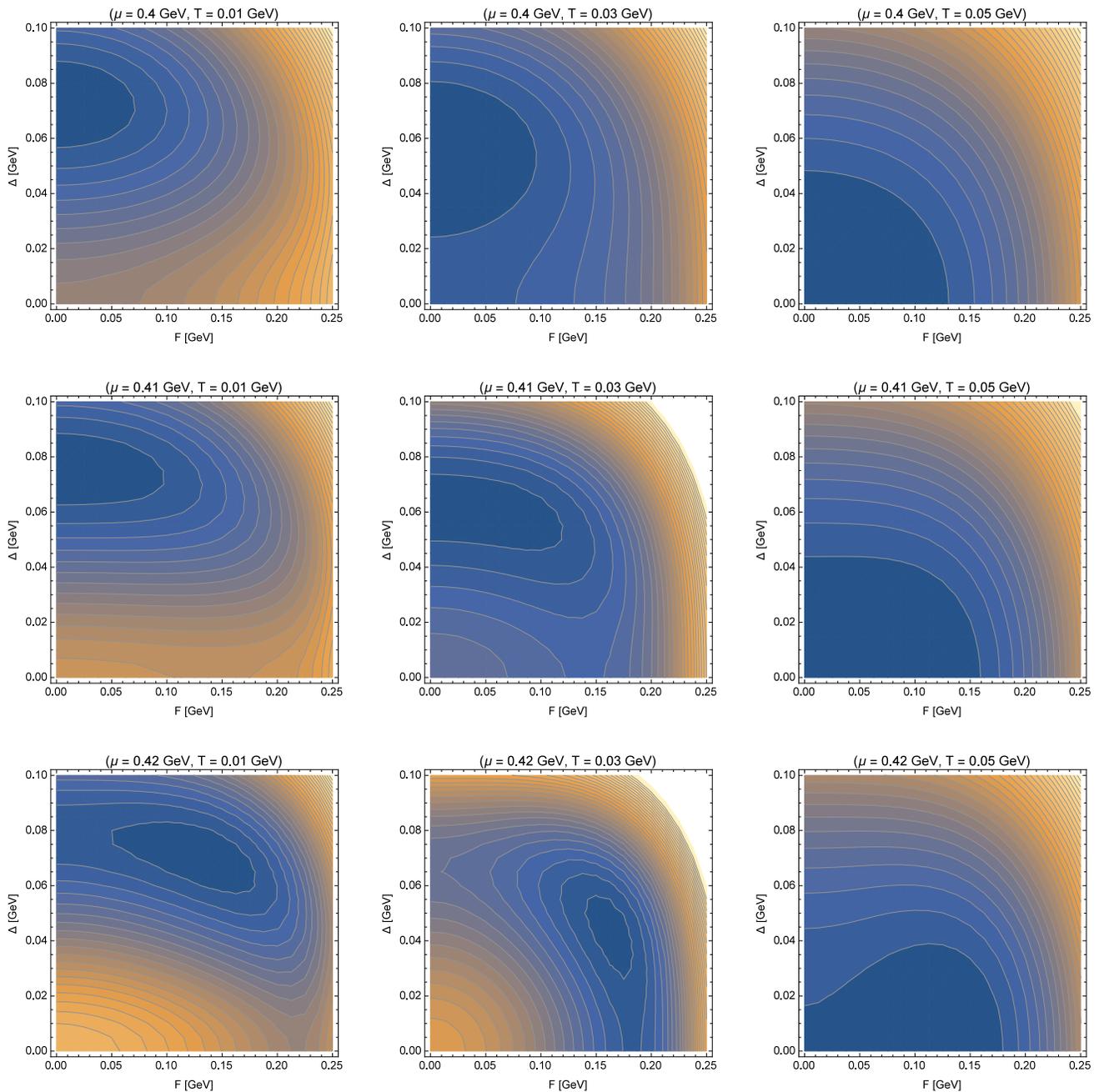}
	\caption{The figures show the contour maps of the thermodynamic potential for model GT2 with 
	several values of the chemical potential, $\mu$ and temperature, $T$.
	The horizontal and vertical axes represent the order parameters for the spin polarization and
	the color superconductivity, respectively.
	The darker color represents lower values of the thermodynamic potential.}
	\label{map2}
\end{figure*}

First we discuss the relationship between the chiral condensed phase and the color superconducting phase
by using model GT0 in Table I, putting to zero the tensor term.
We show numerical results of the thermodynamic potential in Fig \ref{map1}.
The horizontal and vertical axes represent the order parameters for the chiral condensate, $M$, and the color superconducting gap, $\varDelta$, respectively.
The darker color represents lower value of the thermodynamic potential.

At chemical potential $\mu=0.25$ GeV and temperature $T=0.01,\,0.08$ GeV, only the chiral condensed phase is realized.
However, at $\mu=0.25$ GeV and $T=0.12$ GeV, both phases, the chiral condensed phase and the color superconducting phase, disappear.

At $\mu=0.32$ GeV and $T=0.01$ GeV, the chiral condensed phase exists and color superconducting gap does not appear.
But there are two local minima on the horizontal and vertical axes, respectively.
At $\mu=0.32$ GeV and $T=0.03$ GeV, there are also two local minima.
The thermodynamic potential takes about the same value at these points.
At $\mu=0.32$ GeV and $T=0.05$ GeV, there is no condensate.

On the other hand, when $\mu=0.35$ GeV and $T=0.01$ GeV, the color superconducting phase appears but the chiral condensate disappears.
At $\mu=0.35$ GeV and $T=0.02$ GeV, the color superconducting phase is the only phase realized.
Like other cases, in the high temperature region ($\mu=0.35$ GeV and $T=0.05$ GeV), there is no condensate.

These contour plots indicate that the chiral condensed phase and the color superconducting phase do not coexist in our parameter set.
In Ref.~\cite{Blaschke:2003}, however, it has been shown that if one adopts a different parameter set, the two phases may coexist.

It is known that there is an end point where the order of the phase transition between the chiral condensed phase and chiral symmetric phase changes in the phase diagram in the $T$-$\mu$ plane.
In the low chemical potential region, the phase transition is of second order, on the other hand, in the large chemical potential region, it is of the first order.

According to the numerical results, if the chiral condensate is realized, the thermodynamic potential always takes the minimum value on the horizontal axes.
On the other  hand, if the color superconducting gap is realized, the thermodynamic potential always takes the minimum value on the vertical axes.
Thus, we consider that the order of the phase transition between the chiral condensed phase and the color superconducting phase is first order.

\subsection{Spin polarization versus color superconductor}

Next we discuss the relationship between the spin polarization and the color superconductivity
by using model GT2, with $G_T = 2 G_S$ in Table I.
We plot the thermodynamic potential in Fig \ref{map2}.
The horizontal and vertical axes represent the order parameter for the spin polarization and the color superconductivity, respectively.
The darker color represents lower values of the thermodynamic potential.

At chemical potential $\mu=0.4$ GeV and temperature $T=0.01,\,0.03$ GeV, the thermodynamic potential takes the minimum value at $F=0$ and $\varDelta \neq 0$.
Thus the color superconducting phase is realized.
At $\mu=0.4$ GeV and $T=0.05$ GeV, the thermodynamic potential takes the minimum value at the origin.
It means that the simple quark phase is realized.

At $\mu=0.41$ GeV and $T=0.01$ GeV, there is the only a color superconducting gap.
However, at $\mu=0.41$ GeV and $T=0.03$ GeV, the thermodynamic potential takes the minimum value at $F \neq 0$ and $\varDelta \neq 0$.
It indicates that two phases may coexist.
At $\mu=0.41$ GeV and $T=0.05$ GeV, both condensates disappear and the simple quark phase is realized.

At $\mu=0.42$ GeV and $T=0.01,\,0.03$ GeV, the color superconducting gap and the spin polarized condensate coexist.
Then, when $\mu=0.42$ GeV and $T=0.05$ GeV, the color superconducting gap disappears and the spin polarized condensate is realized.
If we set higher temperatures, the thermodynamic potential will take the minimum at the origin, namely, no condensate appear.

According to these contour maps, we consider that the spin polarized phase and the color superconducting phase can coexist in certain conditions, and the order of the phase transition between the color superconducting phase, the spin polarized phase and the coexisting phase is of the second.

\subsection{Phase diagram}

\begin{figure}[]
	\centering
	\includegraphics[width=\columnwidth]{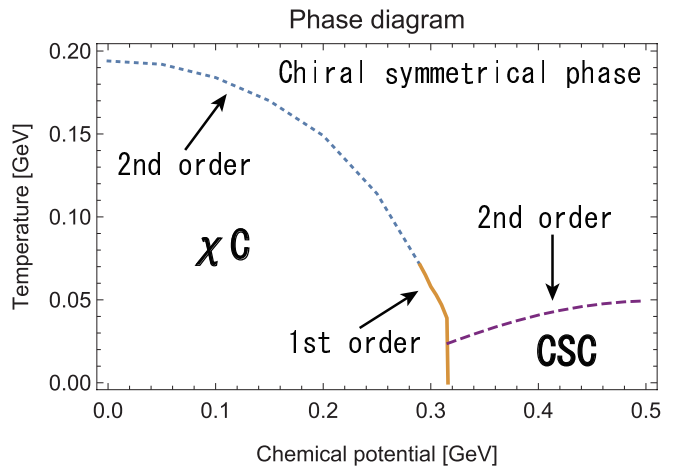}
	\caption{The phase diagram for model GT0 with $G_T = 0$: The horizontal and vertical axes represent 
	chemical potential and temperature, respectively.
	In the figure, $\chi$C and CSC mean the chiral condensed phase and the color
	superconducting phase, respectively.
	``2nd order'' and ``1st order'' mean the order of the phase transition.}
	\label{phase_diagram_0}
\end{figure}

\begin{figure}[]
	\centering
	\includegraphics[width=\columnwidth]{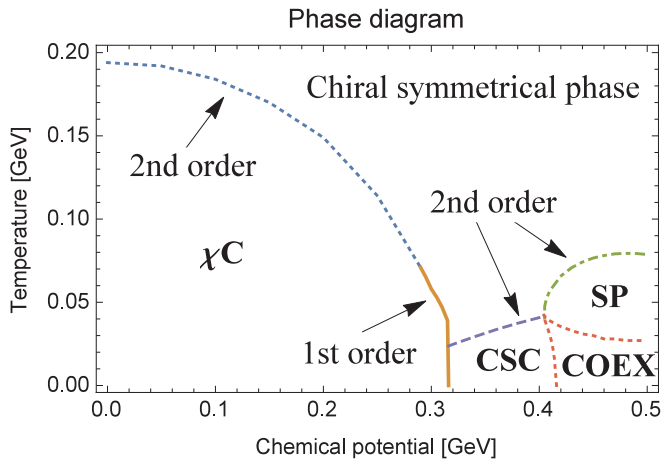}
	\caption{The phase diagram for model GT2 with $G_T = 2 G_S$: The horizontal and vertical axes represent 
	chemical potential and temperature, respectively.
	In the figure, $\chi$C, CSC and SP mean the chiral condensed phase, the color
	superconducting phase and the spin polarized phase, respectively.
	Also, COEX means the coexisting phase with both the spin polarization and the color superconducting gap.
	``2nd order'' and ``1st order'' mean the order of the phase transition.}
	\label{phase_diagram_1}
\end{figure}

Here we show the phase diagram in the $T$-$\mu$ plane in Figs. \ref{phase_diagram_0} and \ref{phase_diagram_1}.
The horizontal and vertical axes represent quark chemical potential and temperature, respectively.

First we show the phase diagram with the chiral condensate and color superconductivity, namely, model GT0.
In Fig \ref{phase_diagram_0} the region under the dotted blue and solid yellow line is the chiral condensed phase.
On the other hand, the region under the purple dashed line is the color superconducting phase.
The terms ``1st order'' and ``2nd order'' mean the order of phase transition, respectively.

Next, using model GT2, in Fig. \ref{phase_diagram_1} the phase diagram with the chiral condensate, color superconductivity and spin polarization is shown.
The left region of the blue dotted and yellow solid lines represents the chiral condensed phase.
The blue dotted and yellow solid lines mean the second- and first-order phase transition between 
the chiral condensed phase and the chiral symmetrical phase, respectively.

The middle region below the violet dashed line is the color superconducting phase.
The color superconducting phase can exist in the low temperature region, and, as chemical potential increases,
the critical temperature for the phase transition between color superconducting and normal phase increases.

The right region of the orange dotted line is the coexisting phase where the color superconducting gap and the spin
polarized condensate coexist.
The critical temperature for the coexisting phase decreases as chemical potential increases.
Note that just above the chemical potential $\mu=0.4$ GeV and at low temperature the color superconducting
phase exists, however, if we increase temperature slightly, we arrive at the coexisting phase.
The region above the coexisting phase, namely, below the green dash-dotted line, is a spin polarized phase.

\begin{figure}[b]
	\centering
	\includegraphics[width=\columnwidth]{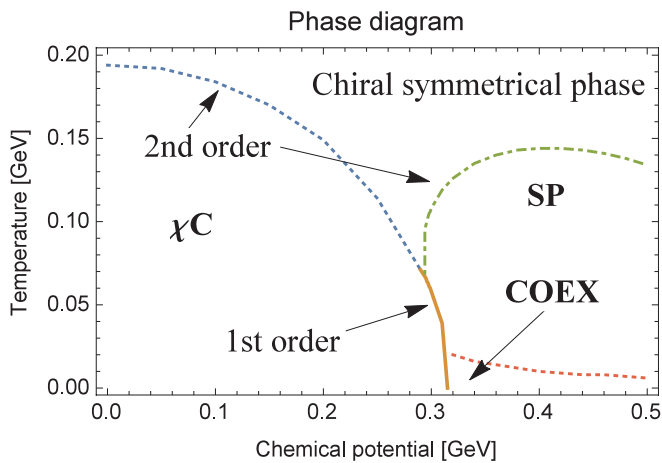}
	\caption{The phase diagram for model GT2.6 with $G_T = 2.6 G_S$: The horizontal and vertical axes represent chemical potential and temperature, respectively.
	In the figure, $\chi$C and SP mean the chiral condensed phase and spin polarized phase, respectively.
	Also, COEX means the coexisting phase with the spin polarization and color superconducting gap.
	``2nd order'' and ``1st order'' mean the order of the phase transition.}
	\label{phase_diagram_2}
\end{figure}

\subsection{Effect of the coupling constant $G_T$}
Finally we investigate the effect of the coupling constant $G_T$ on the thermodynamics.
Here, we have used $G_T$ as a free parameter.
So we give the phase diagram with a new coupling constant: $G_T = 2.6G_S = 14.3 \; \text{GeV}^{-2}$,
identified as model GT2.6 in Table I.
Figure \ref{phase_diagram_2} shows the phase diagram for model GT2.6 in the $T$-$\mu$ plane.
One of the differences between the former phase diagram and the later one is that the color superconducting phase does not appear.
Instead, the coexistence phase is realized after the chiral condensed phase disappears.
Moreover, the spin polarized phase survives at higher temperatures, on the other hand, 
the coexisting phase disappears at lower temperatures than the former phase diagram.

\section{Conclusions and remarks}
\label{sec4}
We have studied the relationship between the chiral condensation, the color superconductivity and the spin polarization at finite density and temperature.
According to the results, in the low chemical potential and temperature region, the chiral condensed phase exists and there is an end point where the order of the phase transition changes.
In the intermediate chemical potential and low temperature region, the color superconducting phase exists.
The chiral condensed and the color superconducting phases do not coexist in our parameter set, however, if we change the values of the coupling strengths and/or three-momentum cutoff parameter, they may coexist.
As is known well, the first order phase transition occurs from the chiral condensed phase to the color superconducting phase in the low temperature region.
When we use model GT2, in the large chemical potential and low temperature region, the color superconducting phase and the spin polarized phase coexist.
However, if we increase temperature, the color superconductivity disappears soon and the only spin polarized phase is realized.
The order of the phase transition between these phases is second order.
At higher temperatures, there are no condensates.
The extension of the spin polarization and color superconducting phase domains depends on the strength of the coupling of the tensor term.

Here we refer to the effect of the coupling constant, $G_T$, to the phase diagram.
We have also examined several values of $G_T$: $G_T = 1.5 G_S,\; 1.75 G_S,\; 1.8 G_S,\; 3 G_S$ and $3.5 G_S$.
When $G_T = 1.5 G_S$, we obtain neither the spin polarized phase nor the coexisting phase, namely,
the phase diagram obtained with this condition is identical to Fig. \ref{phase_diagram_0}.
Thus, if we want these phases to be realized, the value of $G_T$ must be larger than $1.5 G_S$.
When $G_T = 1.75 G_S$ and $1.8 G_S$, we can get phase diagrams qualitatively identical to the one for $G_T = 2 G_S$ (see Fig. \ref{phase_diagram_1}).
When $G_T = 3 G_S$, we obtain a phase diagram qualitatively identical to $G_T = 2.6 G_S$ (see Fig.  \ref{phase_diagram_2}).
If $G_T = 3.5 G_S$, the spin polarized phase is realized at $\mu = 0$, therefore, we consider that the value is too large.

In this paper we have considered the Lagrangian density with flavor $SU(2)$ and color $SU(3)$ symmetry.
However, at the large chemical potential, we should not ignore contributions from strange-quark.
So, our next task is to consider the Lagrangian density with flavor $SU(3)$ symmetry.
In this case a CFL phase may exist.
It is also interesting to consider the effects from an external magnetic field on the spin polarization, although effects from magnetic fields on the QCD phase diagram have been investigated by many researchers \cite{Menezes:2009,Andersen:2016}.
Moreover, in order to describe compact stars, we should consider charge neutrality and $\beta$-equilibrium.
In the present work we have considered $\mu_u = \mu_d = \mu$, where $\mu_u$ and $\mu_d$ are chemical potentials for up- and down-quarks, respectively, and no charge neutrality. 
We expect that the spin polarized phase may generate a strong magnetic field around neutron stars. 
The effects from charge neutrality and $\beta$-equilibrium have been discussed, for example, in Ref. \cite{Menezes:2003}. 
This will be considered in future works.

\section*{ACKNOWLEDGMENTS}
We are grateful to the referee for giving helpful comments.
C.P. thanks Brigitte Hiller for helpful discussions.
Y.T. is partially supported by the Grants-in-Aid of the Scientific Research 
(No. 26400277) from the Ministry of Education, Culture, Sports, Science and 
Technology in Japan. 

\appendix
\appendix
\section{Brief note for charge conjugate matrix and Dirac matrices}

In Sec. \ref{sec2} we transform the Lagrangian density into bilinear form for the quark fields.
Here we show how we can do it.
We use gamma matrices represented by Dirac representation.
The charge conjugate is defined by
\begin{equation}
\bar{\psi}^C := \psi^T C,\quad \psi^C := C \bar{\psi}^T,
\end{equation}
where the charge conjugate matrix is $C := i \gamma^0 \gamma^2$.
We enumerate properties of the $C$ matrix and gamma matrices:
\begin{equation*}
C^\dagger = -C,\quad C^2=-1,
\end{equation*}
and
\begin{equation*}
(\gamma^0)^T = \gamma^0,\quad (\gamma^1)^T = -\gamma^1,\quad (\gamma^2)^T = \gamma^2,\quad
(\gamma^3)^T = -\gamma^3.
\end{equation*}
When the $C$ matrix operates on gamma matrices, we obtain
\begin{equation*}
C \gamma^0 C = \gamma^0,\quad C \gamma^1 C = -\gamma^1,\quad C \gamma^2 C = \gamma^2,\quad 
C \gamma^3 C = -\gamma^3.
\end{equation*}
Using these properties, we can get the following expressions:
\begin{gather}
\bar{\psi}^C \psi^C = \bar{\psi} \psi, \\
\bar{\psi}^C \gamma^0 \psi^C = -\bar{\psi} \gamma^0 \psi \\
\bar{\psi}^C \overleftarrow{\partial}_\mu \gamma^\mu \psi^C = -\bar{\psi} \gamma^\mu \partial_\mu \psi, \\
\bar{\psi}^C \gamma^1 \gamma^2 \psi^C = -\bar{\psi} \gamma^1 \gamma^2 \psi,
\end{gather}
where the last line means $\bar{\psi}^C \varSigma_z \psi^C = -\bar{\psi} \varSigma_z \psi$.


\begin{thebibliography}{00}
\bibitem{Nambu:1961_1}
Y.~Nambu and G.~Jona-Lasinio,
Phys. Rev {\bf 122}, 345 (1961).

\bibitem{Nambu:1961_2}
Y.~Nambu and G.~Jona-Lasinio,
Phys. Rev {\bf 124}, 246 (1961).

\bibitem{Klevansky:1992}
S.~P.~Klevansky,
Rev. Mod. Phys. {\bf 64}, 649 (1992).

\bibitem{Hatsuda:1994}
T.~Hatsuda and T.~Kunihiro,
Phys. Rep. {\bf 247}, 221 (1994).

\bibitem{Alford:2001}
M.~Alford,
Annu. Rev. Nucl. Part. Sci. {\bf 51}, 131 (2001).

\bibitem{Buballa:2002}
M.~Buballa and M.~Oertel,
Nucl. Phys. A {\bf 703}, 770 (2002).

\bibitem{Buballa:2005}
M.~Buballa,
Phys. Rep. {\bf 407}, 205 (2005).

\bibitem{Harding:2006}
A.~K.~Harding and D.~Lai,
Rep. Prog. Phys. {\bf 69}, 2631 (2006).

\bibitem{Tatsumi:2000}
T.~Tatsumi,
Phys. Lett. B {\bf 489}, 280 (2000).

\bibitem{Tatsumi:2003}
T.~Tatsumi, T.~Maruyama, and E.~Nakano,
arXiv:hep-ph/0312351.

\bibitem{Tatsumi:2004}
T.~Tatsumi, T.~Maruyama, and E.~Nakano,
Prog. Theor. Phys. Suppl. No. 153, 190 (2004)

\bibitem{Nakano:2003}
E.~Nakano, T.~Maruyama, and T.~Tatsumi,
Phys. Rev. D {\bf 68}, 105001 (2003).

\bibitem{Maedan:2007}
S.~Maedan,
Prog. Theor. Phys. {\bf 118}, 729 (2007).

\bibitem{Kitazawa:2002}
  M.~Kitazawa, T.~Koide, T.~Kunihiro, and Y.~Nemoto,
  Prog. Theor. Phys. {\bf 108}, 929 (2002).

\bibitem{Bohr:2012}
H.~Bohr, P.~K.~Panda, C.~Provid\^{e}ncia, and J.~da~Provid\^{e}ncia,
Int. J. Mod. Phys. E {\bf 22}, 1350019 (2013).

\bibitem{Tsue:2012}
Y.~Tsue, J.~da~Provid\^{e}ncia, C.~Provid\^{e}ncia, and M.~Yamamura,
Prog. Theor. Phys, {\bf 128}, 507 (2012).


\bibitem{Ferrer:2014}
  E.~J.~Ferrer, V.~de~la~Incera, I.~Portillo, and M.~Quiroz,
  Phys. Rev. D {\bf 89}, 085034 (2014).

\bibitem{Tsue:2013}
Y.~Tsue, J.~da~Provid\^{e}ncia, C.~Provid\^{e}ncia, M.~Yamamura, and H.~Bohr,
Prog. Theor. Exp. Phys. {\bf 2013}, 103D01 (2013).

\bibitem{Tsue:2015_1}
Y.~Tsue, J.~da~Provid\^{e}ncia, C.~Provid\^{e}ncia, M.~Yamamura, and H.~Bohr,
Prog. Theor. Exp. Phys. {\bf 2015}, 103D02 (2015).

\bibitem{Matsuoka:2016}
  H.~Matsuoka, Y.~Tsue, J.~da~Provid\^{e}ncia, C.~Provid\^{e}ncia, M.~Yamamura, and H.~Bohr,
  Prog. Theor. Exp. Phys. {\bf 2016}, 053D02 (2016).

\bibitem{Tsue:2015_2}
Y.~Tsue, J.~da~Provid\^{e}ncia, C.~Provid\^{e}ncia, M.~Yamamura, and H.~Bohr,
Prog. Theor. Exp. Phys. {\bf 2015}, 103D01 (2015).


\bibitem{Bellac}
M.~Le~Bellac,
\textit{Thermal Field Theory} (Cambridge University Press, Cambridge, England,1996).

\bibitem{Blin:1988}
A. Blin, B. Hiller, and M. Schaden, Z. Phys. {\bf 331}, 75 (1988).

\bibitem{Jaminon:1998}
M. Jaminon and E. R. Arriola, Phys. Lett. B {\bf 443}, 33 (1998).

\bibitem{Jaminon:2002}
M. Jaminon, M. C. Ruvio, and C. A. de Sousa, Int. J. Mod. Phys. A {\bf 17}, 4903 (2002).

\bibitem{Battistel:2016}
O. A. Battistel, T. H. Pimenta, and G. Dallabona, Phys. Rev. D {\bf 94}, 085011 (2016).

\bibitem{Berges:1999}
J. Berges and K. Rajagopal, Nucl. Phys. B {\bf 538}, 215 (1999). 


\bibitem{Blaschke:2003}
D.~Blaschke, M.~K.~Volkov, and V.~L.~Yudichev,
Eur. Phys. J. A {\bf 17}, 103 (2003).

\bibitem{Andersen:2016}
J.~O.~Andersen, W.~R.~Naylor and A. Tranberg,
Rev. Mod. Phys. {\bf 88}, 025001 (2016).

\bibitem{Menezes:2009}
D.~P.~Menezes, M.~B.~Pinto, S.~S.~Avancini, A.~P.~Mart\'{i}nez, and C.~Provid\^{e}ncia,
Phys. Rev. C {\bf 79}, 035807 (2009).

\bibitem{Menezes:2003}
D. P. Menezes and C. Provid\^encia, Phys. Rev. C {\bf 68}, 035804 (2003). 



\end{thebibliography}



\end{document}